\documentclass[final]{svjour3}
\usepackage{graphicx}
\usepackage{rotating}
\usepackage{amssymb}
\usepackage{mathptmx}
\makeatletter
\journalname{Journal of Low Temperature Physics}

\usepackage{amsmath}
\usepackage{multicol}
\usepackage{layouts}

\begin{document}

\newcommand{\hdblarrow}{H\makebox[0.9ex][l]{$\downdownarrows$}-}
\title{Model and Measurements of an Optical Stack for Broadband Visible to Near-Infrared Absorption in TiN MKIDs}

\author{K. Kouwenhoven$^{1,2}$ \and I. Elwakil$^1$ \and J. van Wingerden$^3$ \and V. Murugesan$^1$ \and D.J. Thoen$^{2,4}$ \and J.J.A. Baselmans$^{1,2}$ \and P.J. de Visser$^1$}

\institute{
1: SRON Netherlands Institute for Space Research,\\ Niels Bohrweg 4, 2333 CA Leiden, The Netherlands\\
\email{k.kouwenhoven@sron.nl}
\\
\\ 2: Department of Microelectronics, Delft University of Technology,\\ Mekelweg 4, 2628 CD Delft, The Netherlands\\
\\ 3: Else Kooi Laboratory, Delft University of Technology,\\ Feldmannweg 17, 2628 CT Delft, The Netherlands\\
\\ 4: Kavli Institute of Nanoscience, Delft University of Technology,\\ Lorentzweg 1, 2628 CJ Delft, The Netherlands}

\maketitle

\begin{abstract}
Typical materials for optical Microwave Kinetic Inductance Detetectors (MKIDs) are metals with a natural absorption of $\sim$30-50\% in the visible and near-infrared. To reach high absorption efficiencies (90-100\%) the KID must be embedded in an optical stack. We show an optical stack design for a 60 nm TiN film. The optical stack is modeled as sections of transmission lines, where the parameters for each section are related to the optical properties of each layer. We derive the complex permittivity of the TiN film from a spectral ellipsometry measurement. The designed optical stack is optimised for broadband absorption and consists of, from top (illumination side) to bottom: 85 nm $\mathrm{SiO_2}$, 60 nm TiN, 23 nm of $\mathrm{SiO_2}$, and a 100 nm thick Al mirror. We show the modeled absorption and reflection of this stack, which has $>$80\% absorption from 400 nm to 1550 nm and near-unity absorption for 500 nm to 800 nm. We measure transmission and reflection of this stack with a commercial spectrophotometer. The results are in good agreement with the model.


\end{abstract}

\section{Introduction}
\begin{figure}
        \centering
        \includegraphics[width=\linewidth]{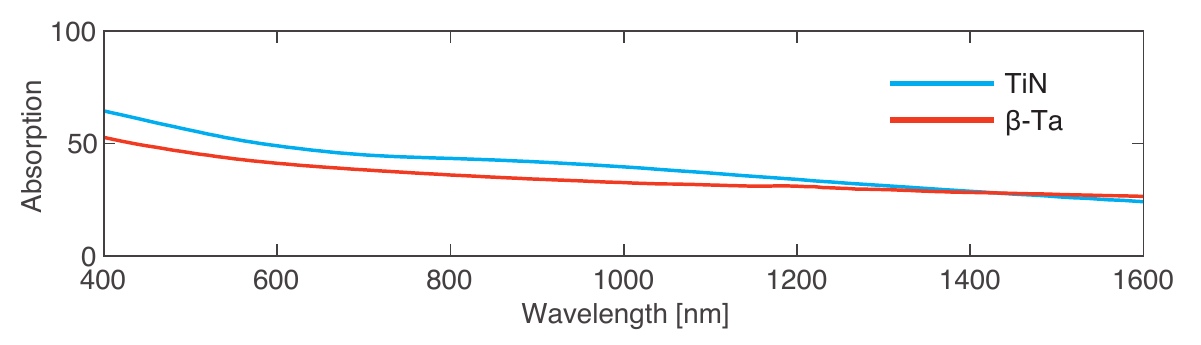}
        \caption{Absorbed power under normal incidence for two high resistivity superconductors, TiN ({\it blue}) and beta phase tantalum ($\mathrm{\beta}$-Ta) ({\it red}). The plotted absorption is calculated from the measured optical constants of both films. Both films are 60 nm thick and deposited on a 350 $\mathrm{\mu}$m polished c-plane sapphire substrate. TiN: $T_c$ = 3.45 K, $R_s$ = 237.2 $\mathrm{\Omega}$. $\beta$-Ta: $T_c$ = 0.99 K, $R_s$ = 38.8 $\mathrm{\Omega}$. (Color figure online.)}
        \label{fig:Absorption_KIDs}
    \end{figure}

There is a wide variation in the superconductors used for Microwave Kinetic Inductance Detectors (MKIDs) \cite{Mazin2020}. Some common materials are Hafnium (Hf), Titanium nitride (TiN), Platinum Silicide (PtSi), and alpha- or beta-phase Tantalum (Ta). Although the properties of these metals differ, they have one thing in common. They suffer from low absorption efficiency in the visible to near-IR wavelength range. For $\beta$-Ta, TiN, PtSi, and Hf absorption is limited to around 50\% in the visible and 30\% in the IR range, as illustrated by Fig.~\ref{fig:Absorption_KIDs} and presented in \cite{Coiffard2020}. To increase the power absorbed in the MKID, we embed it in an optical stack consisting of a mirror (cavity) and one or multiple matching layers. With this approach, unity absorption can be achieved \cite{Rosenberg2004,Lita2008,Dai2019}. However, the bandwidth over which high absorption is achieved is usually limited. Here we show the design of an optical stack for a 60 nm thick TiN layer, based on a transmission line model, that achieves absorption $>$ 80\% over a band from 400 nm to 1550 nm, with near-unity transmission from 500 nm to 800 nm. We verify the model with a reflection and transmission measurement using a commercial spectrophotometer. 

\section{Model}

\begin{figure}
        \centering
        \includegraphics[]{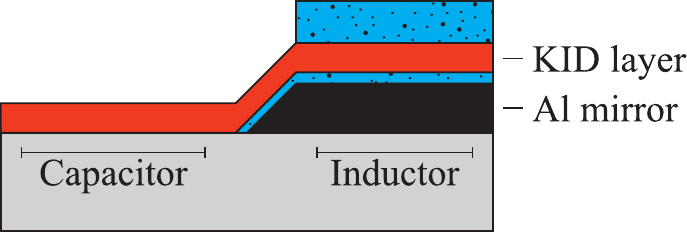}
        \caption{Illustration of the proposed optical stack. The MKID layer ({\it red}) is backed by a 100 nm thick superconducting Al mirror ({\it black}). The stack contains two dielectric layers ({\it blue}), one to create the back-short spacing between the MKID and the superconducting mirror and a second to reduce reflection from the MKID layer. The number of dielectric layers deposited on top of the MKID can be increased to improve the stacks performance when necessary. To limit the added dielectric loss and Two Level System (TLS) noise from the deposited dielectrics, the capacitor should not be embedded in the stack. In this case the MKID is front-side (top) illuminated. (Color figure online.)}
        \label{fig:Concept}
    \end{figure}

The optical stack consists of the superconducting MKID layer (TiN), a superconducting Al mirror, and two dielectric layers, as illustrated in Fig.~\ref{fig:Concept}. The 100 nm thick mirror is deposited on top of the sapphire substrate and has to be made from a superconducting material to maintain the high quality factor of the MKIDs \cite{Dai2019}. The basic version contains two dielectric layers. One layer realizes the spacing between the MKID layer and the mirror. The other layer, deposited on top of the MKID layer, is used to match the impedance of the MKID plus mirror to the impedance of free-space $\eta_0$. The number of layers on top of the MKID can be increased to realize either multi-layer interference or tapered impedance matching structures.

The deposited dielectrics will increase the dielectric loss and Two-Level System (TLS) noise levels in the MKID, especially in regions with a high E-field density. For this reason, the optical stack should be limited to the inductor area.

\subsection{Transmission Line Representation}

    \begin{figure}
        \centering
        \includegraphics[width = 0.8\linewidth]{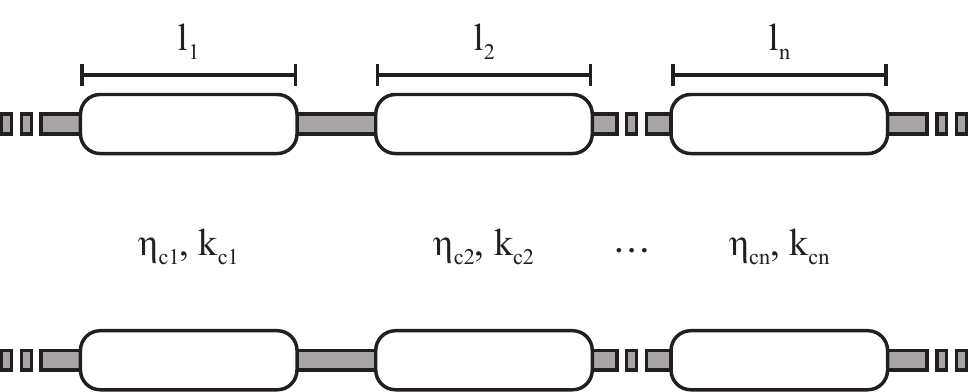}
        \caption{Transmission line representation of the optical stack. Each of the layers, either dielectric or metal, is represented by a section of transmission line with length equal to the height of the layer. The transmission lines are characterized by a complex impedance $\eta_c$ and a complex propagation constant $k_c$. (Color figure online.)}
        \label{fig:TL-line}
    \end{figure}

To model the expected absorption of the optical stack based on the optical properties of each layer, we represent each layer with a section of transmission line as illustrated in Fig.~\ref{fig:TL-line}. The length $l$ of the transmission line is the thickness of the respective layer, and the complex impedance $\eta_c$ and propagation constant $k_c$ are a function of the layer's optical properties. We characterize each layer by its relative permittivity $\Tilde{\epsilon_r}(\omega) = \epsilon'_r(\omega) - i\epsilon''_r(\omega)$, where $\omega$ indicates the frequency dependency of the relative permittivity. For ease of readability, we will omit the frequency dependency of $\Tilde{\epsilon_r}(\omega)$ in the following equations. The transmission line parameters are given by \cite{Pozar2011}

    \begin{equation}
        k_c = \beta_d \sqrt{1-i\tau},
    \end{equation}

    \begin{equation}
        \eta_c = \eta_d \frac{1}{\sqrt{1-i\tau}},
    \end{equation}


where $\beta_d$ and $\eta_d$ are the phase constant and impedance for a plane-wave in a lossless dielectric $\beta_d = \frac{2\pi}{\lambda_0} \sqrt{\epsilon'_r}$ and $\eta_d = \frac{\eta_0}{\sqrt{\epsilon'_r}}$ with $\eta_0$ the free space impedance and $\lambda_0$ the wavelength in free space. The electric loss tangent $\tan\delta_e$ is given by

    \begin{equation}
        \tau = \tan\delta_e = \frac{\epsilon''_r}{\epsilon'_r}.
    \end{equation}

For a perfect conductor $\tau = \infty$ and for a lossless dielectric $\tau = 0$. The ABCD matrix is a useful tool for the analysis of the full optical stack. The ABCD matrix of a transmission line section is \cite{Pozar2011}

    \begin{equation}
        \begin{bmatrix}
        A & B\\
        C & D\\
        \end{bmatrix}
        =
        \begin{bmatrix}
        \cos(k_c l) & \eta_c i\sin(k_c l)\\
        \eta_c^{-1} i\sin(k_c l) & \cos(k_c l)\\
        \end{bmatrix}
        \label{eq:ABCD_transmission}
    \end{equation}

and the matrix of the full stack is then given by the cascade of the ABCD matrices corresponding to the different layers \cite{Pozar2011}

    \begin{equation}
        \begin{bmatrix}
        A & B\\
        C & D\\
        \end{bmatrix}
        =
        \begin{bmatrix}
        A_1 & B_1\\
        C_1 & D_1\\
        \end{bmatrix}
        \begin{bmatrix}
        A_2 & B_2\\
        C_2 & D_2\\
        \end{bmatrix}
        \cdots
        \begin{bmatrix}
        A_n & B_n\\
        C_n & D_n\\
        \end{bmatrix}.
        \label{eq:ABCD cascade}
    \end{equation}

The reflection ($\Gamma$) and transmission ($T$) coefficients of the stack are obtained as \cite{Frickey1994}

    \begin{equation}
        \Gamma = \frac{A Z_{02} + B - C Z_{01}^* Z_{02} - D Z_{01}^*}{A Z_{02} + B + C Z_{01} Z_{02} + D Z_{01}} \label{eq:Reflected}
    \end{equation}

    \begin{equation}
        T = \frac{2\sqrt{\operatorname{Re}(Z_{01})\operatorname{Re}(Z_{02})}}{A Z_{02} + B + C Z_{01} Z_{02} + D Z_{01}},
        \label{eq:Transmitted}
    \end{equation}

where $Z_{01}$ and $Z_{02}$ are the terminations of the transmission line \cite{Frickey1994} as given in Fig.~\ref{fig:TL-line}. For the structure in Fig.~\ref{fig:Concept}, $Z_{01}$ is the wave impedance of free space ($\eta_0 \approx 120\pi$ $\mathrm{\Omega}$) and $Z_{02}$ the wave impedance of the substrate ($\eta_0/\sqrt{\epsilon_{r,sub}}$). The absorbed power fraction of the complete stack is then given by $A = 1 - |\Gamma|^2 - |T|^2$.

\subsection{Spectroscopic Ellipsometry}

We characterize the superconducting films with a room temperature spectroscopic ellipsometry measurement. We measure the amplitude component $\psi$ and phase difference $\Delta$ at multiple angles $\Theta$ ranging from 55$^\circ$ to 80$^\circ$ in steps of 5$^\circ$ (typically centered around the Brewster's angle). We assume that the films are isotropic, homogeneous and that the measured reflection is dominated by the first reflection at the air-TiN interface. In this case, the film can be considered infinitely thick. Under these assumptions, the complex permittivity of the film is retrieved as

    \begin{equation}
        \Tilde{\epsilon}(\omega) = \sin^2(\Theta)\left[ 1 + \tan^2(\Theta) \left( \frac{1-\rho}{1+\rho} \right)^2 \right]
    \end{equation}

with $\rho = \tan{(\psi)}\exp{(i\Delta)}$ and $\Tilde{\epsilon_r} = \epsilon'_r - i\epsilon''_r$. The complex permittivity $\Tilde{\epsilon_r}$ is related to the complex refractive index as

    \begin{equation}
        \Tilde{\epsilon_r} = (n+ik)^2.
    \end{equation}

\begin{figure}
        \centering
        \includegraphics[width=\linewidth]{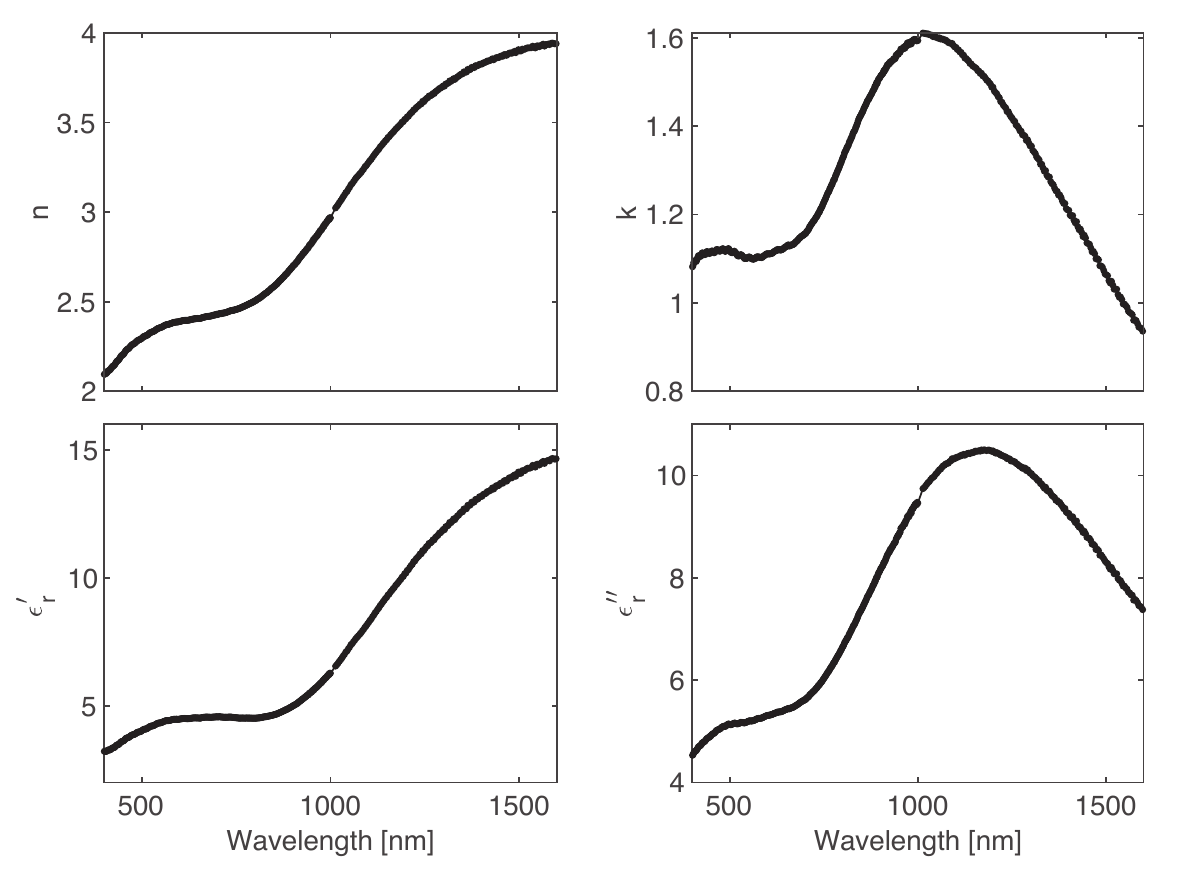}
        \caption{Optical constants (n,k) and relative permittivity for TiN. These values are derived from a multi-angle spectroscopic ellipsometry measurement of a 60 nm thick TiN film deposited on a 350 $\mathrm{\mu}$m polished c-plane sapphire substrate. Individual measurements points are plotted, showing a gap around 1000 nm where the set-up switches between two different detectors. Measured from 191.35 nm to 1688.0 nm in 710 non-equidistant measurement points.}
        \label{fig:nk_data}
    \end{figure}

The derived complex permittivity for a 60 nm thick TiN film, deposited on a sapphire substrate, is given in Fig.~\ref{fig:nk_data}. The TiN film has a critical temperature ($T_c$) of 3.45 K and a sheet resistance ($R_s$) of 237.2 $\mathrm{\Omega}$. 

\subsection{Optimization for Broadband Absorption}
 
For high absorption at single wavelengths, the optimization is easy. The thicknesses of both dielectrics, see Fig.~\ref{fig:Concept}, should be chosen as $\lambda_d/4$ (quarter wavelength) with $\lambda_d$ the wavelength in the dielectric. For absorption over a large wavelength range, however, the optimal thicknesses can no longer be calculated analytically since the optical constants of the metal layer vary strongly with wavelength, as indicated by Fig.~\ref{fig:nk_data}. Therefore we parameterize the optical stack with respect to the layer thicknesses and calculate the achieved absorption for multiple thickness combinations. We optimize for maximum average absorption over a defined wavelength range, in this case, 400 nm to 1550 nm. The optimal layer combination for a 60 nm TiN film consists of, from top (illumination side) to bottom: 85 nm $\mathrm{SiO_2}$, 60 nm of unpatterned TiN, 23 nm of $\mathrm{SiO_2}$, and a 100 nm thick Al mirror. $\mathrm{SiO_2}$ is chosen since it is transparent (lossless) for visible light and has a relative permittivity close to the desired value for a $\lambda/4$ matching layer, $\epsilon_{r,\lambda/4} = \sqrt{\epsilon_{r,1} \epsilon_{r,2}}$ with $\epsilon_{r,1}$ and $\epsilon_{r,2}$ the relative permittivity of the two media surrounding the matching layer. The material properties for $\mathrm{SiO_2}$ and Al are taken from \cite{Rakic1995} and \cite{Marcos2016}.

The calculated absorbed and reflected power fractions for this stack are given in Fig. \ref{fig:results}. The given absorption is for the entire structure and contains the power absorbed by the Al mirror in addition to the power absorbed in the TiN MKID. The transmission line model presented here can be used to extract the power absorbed per layer, separating the power absorbed in the TiN layer from the power lost in the dielectric and Al layers. The model shows that of the power absorbed by the stack, 1-5\% is absorbed in the Al mirror with a $\sim$1\% absorption in the two dielectric layers for shorter wavelengths, see Fig.~\ref{fig:results}. The authors of \cite{Verma2020} quote an absorption of 3\% in the metallic mirror at 1550 nm, similar to the results presented here. For a single wavelength application an optimal stack can be analytically designed for 100\% absorption. However, the power absorbed in the actual detector is limited to $\sim$95-98\% by absorption in the dielectric and Al layers.

\section{Measurement}

    \begin{figure}
        \centering
        \includegraphics[width=\linewidth]{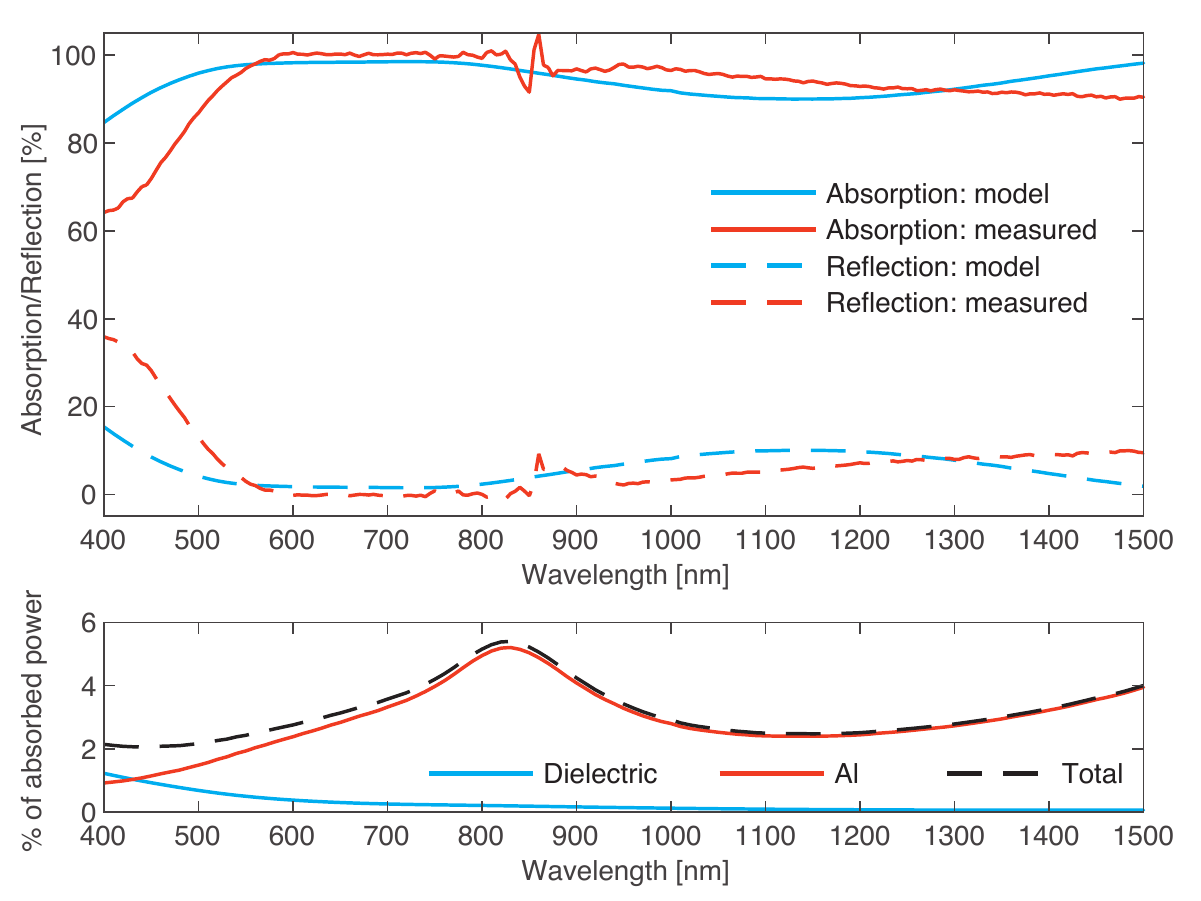}
        \caption{\textbf{Top:} Model ({\it blue}) and measurement ({\it red}) of the absorption ({\it solid}) and reflection ({\it dashed}), of a stack consisting of, from top (illumination side) to bottom: 85 nm layer of $\mathrm{SiO_2}$, a 60 nm thick unpatterned TiN film, a 23 nm $\mathrm{SiO_2}$ layer and a 100 nm thick Al mirror deposited on a 350 $\mathrm{\mu}$m polished c-plane sapphire substrate.The absorbed power is derived from the measured reflection and transmission. The measurement is performed with a PerkinElmer Lambda 1050+ spectrophotometer equipped with an integrating sphere. \textbf{Bottom:} Power lost per material defined as the fraction of total absorbed power. The material properties for $\mathrm{SiO_2}$ and Al are taken from \cite{Rakic1995} and \cite{Marcos2016}. (Color figure online.)}
        \label{fig:results}
    \end{figure}

We measure the reflectance and transmittance from 400 nm to 1550 nm with a commercial spectrophotometer (PerkinElmer Lambda 1050+) equipped with an integrating sphere (IS), with the sample mounted at the front (transmission) or back (reflection) port of the IS. The measurement is self-calibrating through spectralon (white disk) reference measurements. At around 900 nm, the spectrophotometer switches gratings, which causes a calibration error resulting in fluctuations around the 900 nm point. The measured reflection and derived absorption are given in Fig.~\ref{fig:results}, which shows a good agreement between the model and measurement. The transmitted power is omitted from this figure since it is 0\% over the entire wavelength range for both the model and measurement due to the 100 nm thick Al mirror. The $SiO_2$ layers are sputter deposited from a quartz target.

The disagreement between model and measurement can be explained by a relatively small change in optical properties or a deviation in the layer thicknesses. We did not measure the layer thicknesses of the fabricated sample. These changes can explain the larger deviation at shorter wavelengths in Fig.~\ref{fig:results} as they can create destructive interference close to the wavelength range of interest.

The reflection and transmission measurement is performed at room temperature while the operating temperature of MKIDs is around 100 mK. The change in temperature can have an effect on the optical constants of the metallic layers. To study the temperature dependence of these films, the optical constants of metallic layers can be modelled as a combination of a Drude term and a set of Lorentz oscillators, as is done for TiN in \cite{Patsalas2001} and for Al in \cite{Rakic1995}. The Drude term scales with the conductivity of the sample $\sigma$ and is thus temperature dependent. We assume the interband transitions, modelled with the Lorentz oscillator model, are temperature independent. We measure the temperature dependency of $\sigma$ with a four-probe DC structure during a cool-down as the Residual-Resistance Ratio (RRR). For TiN, RRR is $\sim$1 and for Al it is $\sim$3, indicating that the optical constants of TiN should be temperature independent. Furthermore, the analysis in \cite{Rakic1995} and \cite{Patsalas2001} show that the optical constants of both the Al and TiN films are dominated by the temperature independent interband transitions.

\section{Conclusion and Outlook}

We have designed an optical stack consisting of, from top (illumination side) to bottom: 85 nm $\mathrm{SiO_2}$, 60 nm TiN (MKID inductor), 23 nm $\mathrm{SiO_2}$ and a 100 nm thick Al mirror optimized for broadband absorption in the visible to near-infrared. As presented in Fig.~\ref{fig:results}, the initial results are promising, showing near-unity absorption from 600 nm to 800 nm and overall high absorption in the 400 nm to 1550 nm band. For the next iteration, we will measure both the layer thickness and optical constants of each deposited layer, to get a better agreement between model and measurement. The final goal is to measure the absorption efficiency of a MKID with its inductor embedded in an optical stack. In addition to the analysis in this paper, this requires an understanding of the effect of temperature on the optical properties of each layer, as well as the effect of a patterned MKID layer (inductor lines).

\begin{acknowledgements}
We acknowledge Martijn Tijssen and Olindo Isabella for their support with the ellipsometry and reflection/transmission measurements. This work is financially supported by the Netherlands Organisation for Scientific Research NWO (Projectruimte 680-91-127 and Veni Grant No. 639.041.750). JJAB was supported by the European Research Council ERC (Consolidator Grant No. 648135 MOSAIC). The datasets generated during and/or analysed during the current study are available in the ZENODO repository, https://doi.org/10.5281/zenodo.5549994.
\end{acknowledgements}


\end{document}